\documentclass[fleqn,usenatbib]{mnras}

\usepackage{newtxtext,newtxmath}

\usepackage[T1]{fontenc}

\DeclareRobustCommand{\VAN}[3]{#2}
\let\VANthebibliography\thebibliography
\def\thebibliography{\DeclareRobustCommand{\VAN}[3]{##3}\VANthebibliography}

\usepackage{graphicx}	
\usepackage{amsmath}	

\usepackage{svg}
\usepackage{longtable}
\usepackage{booktabs}
\usepackage{lscape}
\usepackage{makecell}
\usepackage{bm}
\usepackage{caption}
\usepackage{subcaption}

\newcommand{\scrutinisedcount}{1168}
\newcommand{\totcount}{134}
\newcommand{\energycount}{13}

\title[SNAD catalogue of M-dwarf flares from ZTF]{SNAD catalogue of M-dwarf flares from the Zwicky Transient Facility}

\author[Voloshina, Lavrukhina et al.]{
A.~S.~Voloshina, $^{1}$\thanks{A.~S.~Voloshina and A.~D.~Lavrukhina contributed equally}
A.~D.~Lavrukhina, $^{2,3}$\thanks{Email: lavrukhina.ad@gmail.com}
M.~V.~Pruzhinskaya, $^{1}$
K.~L.~Malanchev, $^{4,5}$
E.~E.~O.~Ishida, $^{1}$
\newauthor{
V.~V.~Krushinsky, $^{6}$
P.~D.~Aleo, $^{5,7}$
E.~Gangler, $^{1}$
M.~V.~Kornilov, $^{2,8}$
V.~S.~Korolev, $^{9}$
E.~Russeil, $^{1}$}
\newauthor{
T.~A.~Semenikhin, $^{2,3}$
S.~Sreejith, $^{10}$
and A.~A.~Volnova $^{11}$
(The SNAD team)}
\\
$^{1}$Universit\'e Clermont Auvergne, CNRS/IN2P3, LPCA, F-63000 Clermont-Ferrand, France\\
$^{2}$Lomonosov Moscow State University, Sternberg Astronomical Institute, Universitetsky pr.~13, Moscow, 119234, Russia\\
$^{3}$Lomonosov Moscow State University, Faculty of Space Research, Leninsky Gori 1 bld. 52, Moscow 119234, Russia\\
$^{4}$McWilliams Center for Cosmology \& Astrophysics, Department of Physics, Carnegie Mellon University, Pittsburgh, PA 15213, USA\\
$^{5}$Department of Astronomy, University of Illinois at Urbana-Champaign, 1002 West Green Street, Urbana, IL 61801, USA\\
$^{6}$Laboratory of Astrochemical Research, Ural Federal University, Ekaterinburg, ul. Mira d. 19, Yekaterinburg, 620002, Russia\\
$^{7}$Center for Astrophysical Surveys, National Center for Supercomputing Applications, Urbana, IL, 61801, USA\\
$^{8}$National Research University Higher School of Economics, 21/4 Staraya Basmannaya Ulitsa, Moscow, 105066, Russia\\
$^{9}$Independent researcher\\
$^{10}$Physics department, University of Surrey, Stag Hill, Guildford GU2 7XH\\
$^{11}$Space Research Institute of the Russian Academy of Sciences (IKI), 84/32 Profsoyuznaya Street, Moscow, 117997, Russia
}

\date{Accepted XXX. Received YYY; in original form ZZZ}

\pubyear{2024}

\begin{document}
\label{firstpage}
\pagerange{\pageref{firstpage}--\pageref{lastpage}}
\maketitle 
\begin{abstract}
Most of the stars in the Universe are M spectral class dwarfs, which are known to be the source of bright and frequent stellar flares. In this paper, we propose new approaches to discover M-dwarf flares in ground-based photometric surveys. We employ two approaches: a modification of a traditional method of parametric fit search and a machine learning algorithm based on active anomaly detection. The algorithms are applied to Zwicky Transient Facility (ZTF) data release 8, which includes the data from the ZTF high-cadence survey, allowing us to reveal flares lasting from minutes to hours. We analyze over 35 million ZTF light curves and visually scrutinize  \scrutinisedcount{} candidates suggested by the algorithms to filter out artifacts, occultations of a star by an asteroid, and other types of known variable objects. The result of this analysis is the largest catalogue of ZTF flaring stars to date, representing \totcount{} flares with amplitudes ranging from -0.2 to -4.6 magnitudes, including repeated flares. Using Pan-STARRS DR2 colors, we assign a spectral subclass to each object in the sample. For \energycount{} flares with well-sampled light curves and available geometric distances from Gaia DR3, we estimate the bolometric energy. This research shows that the proposed methods combined with the ZTF’s cadence strategy are suitable for identifying M-dwarf flares and other fast transients, allowing for the extraction of significant astrophysical information from their light curves.
\end{abstract}

\begin{keywords}
stars: flare -- stars: late-type -- stars: activity -- surveys -- methods: data analysis
\end{keywords}



\section{Introduction}
M-dwarf stars make up the vast majority of stars in our galaxy.
As low-mass, fully convective stars, they exhibit frequent flaring events caused by powerful magnetic reconnection processes in their atmospheres~\citep{GershbergPikelner}.
The study of M-dwarf flares provided key insight into stellar magnetism, high-energy phenomena, and the impacts on potential planets orbiting these stars.
However, many fundamental properties of M-dwarf flares remain poorly constrained, including their occurrence rates, energy output, and relationship with stellar properties, such as age and metallicity (see for a review \citealt{2024arXiv240207885K}).

Multiple time-domain optical surveys have been utilised for systematic M-dwarf flare search projects. Space-based missions such as Kepler~\citep{2010Sci...327..977B} and Transiting Exoplanet Survey Satellite (TESS;~\citealt{2014SPIE.9143E..20R}) offer high regular cadence and very precise relative photometry, making their data excellent sources of stellar flares, especially those of small amplitude. \cite{yang2019_kepler_flares} discovered approximately $1.6\times10^5$ flares on around 3400 stars in the Kepler data. \cite{tess_dr1_flares} found 8695 flares in the first TESS data release, while \cite{tess_3year_flares} refined this number to roughly $1.4\times10^5$ flares over three years of TESS data. 
While these space missions have revealed a vast number of stellar flares with good completeness, ground-based surveys could complement them. For instance, the Zwicky Transient Facility (ZTF; \citealt{2019PASP..131a8002B}) covers sky areas that Kepler did not observe, with a bigger survey depth than TESS.

The recent advent of wide-field time-domain surveys provided new opportunities to build large statistical samples of stellar flares across a range of spectral types.
Numerous systematic stellar-flare searches were performed with different ground-based surveys, occasionally observing high-amplitude flares.
For example, \citet{sdss_2mass_flares} analysed hundreds of flares from SDSS and 2MASS time-domain surveys with the maximum $u$-band change in magnitude $|\Delta M|$ $\sim4.5$~mag.
The ASAS-SN M-dwarf flare catalogue contains 62 flares with a maximum $|\Delta M|$ being $\sim 2$~mag in the $V$-band \citep{asas-sn_flares}.
\citet{DECam_flares} found 96 flares from 80 stars in the DECam data, with a maximum change in magnitude of $\sim1.8$~mag.
\citet{tomoe-gozen_flares} revealed 22 fast stellar flares from a one-second-cadence survey performed by the Tomo-e Gozen project at the Kiso Schmidt telescope in Japan, among which the largest change in magnitude was of $\sim3.25$~mag.
\citet{tmts_flares} presented a catalogue of 132 flares from 125 stars with $|\Delta M|$ up to $\sim3.1$\,mag found in two-year data of the Tsinghua University-Ma Huateng Telescopes Survey.

In this work, we use photometric data from the Zwicky Transient Facility survey, to detect transient events.
ZTF runs multiple surveys, including a high-cadence survey, which provides a unique data set of minute-scale cadence with $\sim 21.5$ limiting magnitude \citep{Kupfer_2021}.
This makes ZTF data releases a source of well-sampled M-dwarf flare light curves occurring in a few hundred parsecs from the Sun.
\citet{2023arXiv231117862C} used ZTF high-cadence survey for a systematic search for gravitational self-lensing binaries and presented 19 candidates.
However, according to the authors, most of these candidates are likely to be stellar flares, making this data set the largest ZTF stellar flare catalogue previously published.
Additionally, the SNAD team has discovered few stellar flares in ZTF data releases using machine learning anomaly detection pipelines \citep{malanchev2020,2023A&A...672A.111P,Volnova2023}.

Various algorithms have been developed for detecting stellar flares in both sparsely sampled and continuously sampled data. Traditional flare detection studies often relied on detrending light curves and using outlier detection heuristics to identify flare events. These methods focused on preprocessing the data to remove trends and then identifying unusual data points which deviate significantly from expected patterns (e.~g., \citealt{2012ApJ...754....4O,2016ApJ...829...23D,tess_dr1_flares}). Another approach is parametric fitting, which models the light curves of flares using predefined mathematical functions to identify their characteristics~\citep{2019AJ....158..119L}. However, the diversity of flare shapes and the data volumes of ongoing wide-field surveys encourage the community to use machine learning techniques. For instance, \citet{2020AJ....160..219F} developed a convolutional neural network (CNN), \texttt{stella}, specifically trained to find flares in TESS short-cadence data. Another machine-learning algorithm presented by \citet{2018A&A...616A.163V} identified flares by analysing light curves and has been successfully applied to stars such as TRAPPIST-1 and KIC~1722506. The current paper explores parametric fitting and machine learning methodologies, which leads to the detection of a heterogeneous set of flaring stars. Since not all the candidates selected with these methods are stellar flares, we also conduct the dedicated one-by-one analysis to filter out bogus detections and objects of a different astrophysical origin.

This paper provides a catalogue of \totcount{} flaring stars detected using two different methodologies.
We calculate astrophysical properties of stars, such as spectral class and interstellar reddening and analyse well-sampled light curves to estimate the total energy, amplitude, and timescale of the flares.



\section{Data}
\label{section:data}
The Zwicky Transient Facility is a wide-field astronomical survey of the entire northern sky, conducted with the 48-inch Schmidt-type Samuel Oschin Telescope at Palomar Observatory~\citep{2019PASP..131a8002B}. During phase I (Feb 2018 -- Sept 2020), ZTF performed a three-day cadence survey of the visible northern sky and one-day for the Galactic plane.
During phase II (Oct 2020 -- Sept 2023), 50\% of the ZTF camera time is dedicated to a two-day cadence public survey in $g$ and $r$ bands. Data from the public survey are released on a bi-monthly schedule as data releases (DRs). In addition, \citealt{Kupfer_2021} conducted a dedicated high-cadence Galactic plane survey with a cadence of 40 seconds. 

In this work, we analyze the private and public data from ZTF DR8 (March~17, 2018 -- September~3, 2021) as target data sets for searching for red dwarf flares. We use only "clear" (\texttt{catflags} = 0) $r$- and $g$-band observations. ZTF DRs provide unique objects in each passband and observational field, so our dataset may have multiple light curves associated with a single source. In order to conduct further astrophysical analysis of the resulting candidates, we use data from additional catalogues: \textit{riz} magnitudes from Pan-STARRS DR2 \citep{2016Pan-starrs1,2018AAS...23143601F}, geometric distances from Gaia EDR3 \citep{2016A&A...595A...1G,2021A&A...649A...1G,Gaia_EDR3_dist_par}, interstellar extinction from ``Bayestar19'' \citep{Green_2019} and \citealt{2011ApJ...737..103S} extinction maps.


\section{Methods}
\label{section:methods}

In this paper we use two distinct approaches of M-dwarf flare identification. The first method is based on parametric search of flaring light curves in high-cadence subsample of ZTF Data Release 8 (DR8). The second method is based on an active anomaly detection approach applied to the full light curves of the entire ZTF DR8. 
We describe both methods in detail below.

\subsection{Parametric fit search}
As part of the ZTF survey, several different observational campaigns were conducted, including high-cadence survey~\citep{Kupfer_2021}. Unfortunately, bulk-downloadable ZTF DRs do not maintain the connection between each individual observation and the specific campaign to which this observation belongs.
For this reason, it was necessary to develop a method for extracting high-cadence data from an entire DR.

The aim was to chunk light curves to form a dataset of intra-day 
light curves having: 1) enough observations for further analysis, 2) high cadence, and 3) covering the time interval typical for flare duration. 
To achieve the first two conditions, we set the minimum number of observations to 5 and the maximum delay between two consecutive observations to 30 minutes. 
As for the minimum duration of these partial light curves, we produce two samples, one with minimum duration of 2 hours (long-duration sample) and another with 30 minutes (short-duration sample).
Such an approach allows us to detect both, short and long flares. 

Along with the imposed conditions to the cadence and duration, all light curves were filtered according to observed source variability. We consider an object variable if a test based on 1-dimensional  reduced $\chi^2$ statistics rejects the hypothesis concerning its non-variability~\citep{Sokolovsky2017}:
\begin{equation}
\label{reduced_chi2}
    \text{reduced}~\chi^2 \equiv \frac1{N-d} \sum_i\left(\frac{m_i - \bar{m}}{\delta_i}\right)^2,
\end{equation}
where $m_i$ is the observed magnitude and $\delta_i$ its observational error, $\bar{m}$ is the  weighted mean magnitude, $N$ is the number of observations and $d$ the number of model parameters. 
The final dataset consists of light curves with a value of the reduced $\chi^2$ statistics greater than 11.
The total number of intra-day $r$-band light curve chunks in the long-duration sample is 4\,027\,686, and in the short-duration one is 10\,351\,985.

The parametric search method is based on light-curve fitting with an analytical function and subsequently selecting well-fit objects.
For this purpose we adopt a semi-phenomenological model of flux evolution, $f(t)$, from \cite{Mendoza}:

\begin{align}\label{eq:mendoza}
    &f(t) = f^* + \frac{\sqrt{\pi} A C}{2} \times \big(F_1 \, h(t, B, C, D_1) + F_2 \, h(t, B, C, D_2)\big)\,,
    \\
    &h(t, B, C, D) = \exp{(\alpha C D)} \times \text{erfc} (\alpha)\,,
    \\
    &\alpha(t, C, D) = \frac{B - t}{C} + \frac12 C D,
\end{align}
where $f^*$ is stellar (quiescence) flux density, $t$ is time, $A$ is the normalizing factor, $B$ is the reference time, $C$ is the Gaussian heating timescale, $D_1$ is the inverse of the rapid cooling phase timescale, $D_2$ is the inverse of the slow cooling phase
timescale, $F_2 \equiv (1 - F_1) \cdot \exp{(\frac{C^2}{4}(D^2_1 - D^2_2))}$, where $F_1$ and $F_2$ describe the relative importance of the exponential cooling terms, and $\mathrm{erfc}$ is the complimentary error function.
Note that $A$ corresponds to the value used by \cite{Mendoza} multiplied by $\exp{(B^2/C^2 - D_1^2 C^2 / 4)}$.
We also change the form of the equation by introducing the dimensionless function $\alpha(t,C,D)$ for better readability and robustness of the fit\footnote{\cite{Mendoza} used the term $B/C$, which could be considered ill formed, since $B$ has the dimensionality of the date, while $C$ has the dimensionality of the time interval}.

A Python function for fitting a light curve was implemented within the \texttt{light-curve}\footnote{\url{https://github.com/light-curve/light-curve-python}} feature extraction library~\citep{light_curve}.
This function chooses the optimal values of parameters $A$, $B$, $C$, $D_1$, $D_2$, $F_2$ and $f^*$ using least-squares fits provided by \texttt{iminuit}~\citep{iminuit}.
For better performance of the least-squares fitting, we used best-fit coefficients from the Markov Chain Monte Carlo analysis, presented in~\cite{Mendoza}, as initial values.
The mean value of the $3\sigma$-clipped flux is used as the initial value of stellar flux $f^*$.
The fit quality is evaluated using reduced $\chi^2$ statistics~(Eq.~\ref{reduced_chi2}) with $d = 7$.

The next filtering step is to distinguish flares with enough points in the flare from ones with only a few points.
For this reason, we use the \texttt{OtsuSplit} feature of the \texttt{light-curve} package~\citep{OtsuFeature}. 
This feature uses a magnitude threshold to distinguish faint and bright subsamples of a light curve based on maximization of interclass variance.
To filter out candidates, we use $\texttt{lower-to-all} > 0.25$ (ratio of the number of points in the bright subsample to the number of points in the faint one) obtained for each object based on the determined threshold.
However, this method does not guarantee perfect separation of the ``flaring'' part of the light curve from the ``plateau'', so some candidates with few-point flares are still present in the final sample. The total number of candidates obtained with this procedure is 308.

\subsection{Machine learning method}

Active anomaly detection represents a family of machine learning techniques which sequentially uses expert feedback to fine-tune an initially standard unsupervised algorithm to a particular definition of scientifically interesting anomaly. In the implementation used in this work, we employ the Active Anomaly Discovery (AAD) algorithm developed by \citet{Das2018} in the form used by \citet{Ishida2019}. 

\begin{figure*}
\centering
	\includegraphics[width=180mm,scale=1.0]{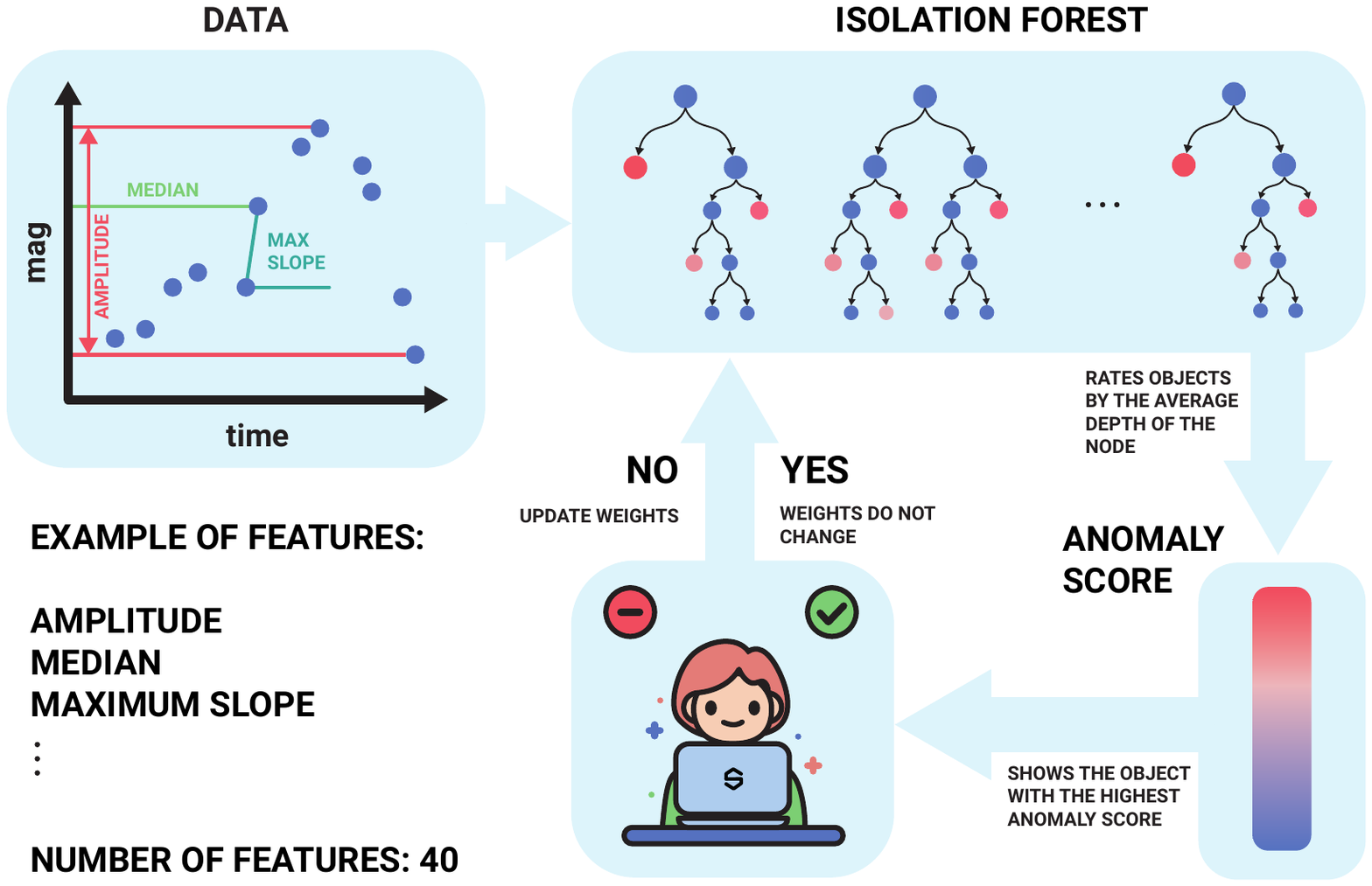}
    \caption{Flowchart of the active machine learning method used to discover M-dwarf flares.}
    \label{fig:AAD_visualisation}
\end{figure*}

The algorithm starts from a traditional isolation forest \citep[IF, ][]{liu2008isolation}. 
It is based on the hypothesis that  objects in under-dense regions of the parameter space (statistical anomalies) are more rapidly isolated from the bulk of the data than nominal ones. In the first step of AAD, a traditional IF is built and the object with the highest anomaly score is shown to a human expert, who is required to provide a binary answer (``YES''/``NO'') to the question: is this anomaly scientifically interesting to you? The expert makes decisions based on both light curve behavior and auxiliary data, such as original scientific images and catalogue cross-matches (see Section~\ref{section:sample-selection} for the details). If the answer is YES, the algorithm will show the second object with the highest anomaly score and pose the same question. Alternatively, if the expert answers NO, the algorithm recalculates the weights corresponding to the decision path which lead to that anomaly score. This modification is applied to the entire forest, the scores are recalculated, and the new object with the highest anomaly score is shown to the expert. After a few iterations, this procedure results in a personalized model which has a lower probability to give high anomaly scores to objects which are not in the expert's main interest \citep[full mathematical description is provided in][]{Das2018}. The flowchart that demonstrates the main steps of the AAD approach is presented in Fig.~\ref{fig:AAD_visualisation}.

Since the algorithm can be adapted to the opinion of the expert, it can be used for a targeted search for objects of a given type (see ~\citealt{2023A&A...672A.111P}). Therefore, in this analysis, a human expert considers only M-dwarf flare candidates as anomalies; all other objects proposed by the algorithm are rejected by the expert as nominals.

The minimum number of observations per light curve was selected to be 300. To avoid nonvariable objects, we select $\chi2 > 3$  (Eq.~\ref{reduced_chi2}, $d=1$). This selection resulted in 21 469 857 light curves.

We run AAD on the obtained data set and visually inspect 860 objects. Among those, we responded ``YES'' to 35 objects.


\section{Sample selection}
\label{section:sample-selection}

Visual inspection by human experts is a part of the sample selection process for both methods: as a final check of the parametric fit outputs, and at each iteration step of the AAD algorithm. All candidates were inspected using the SNAD ZTF Viewer\footnote{\url{https://ztf.snad.space/}}, a special web interface developed by the SNAD team to facilitate expert analysis~\citep{ztf_viewer}. For each object, the Viewer displays its multi-colour light curves and enables easy access to the individual exposure images and to the Aladin Sky Atlas~\citep{2000A&AS..143...33B,2014ASPC..485..277B}. Moreover, it provides cross-matches with various catalogues of stars and transients, including \textsc{SIMBAD} and VizieR databases~\citep{2000A&AS..143....9W}, AAVSO VSX~\citep{2006SASS...25...47W}, Pan-STARRS DR2~\citep{2020ApJS..251....7F}, Gaia DR3~\citep{2023A&A...674A...1G}, and ZTF alert brokers\footnote{\url{https://alerce.science/}}\textsuperscript{,}\footnote{\url{https://antares.noirlab.edu/}}\textsuperscript{,}\footnote{\url{https://fink-portal.org/}}. 

\begin{figure*}
\centering
	\includegraphics[width=180mm,scale=1.0]{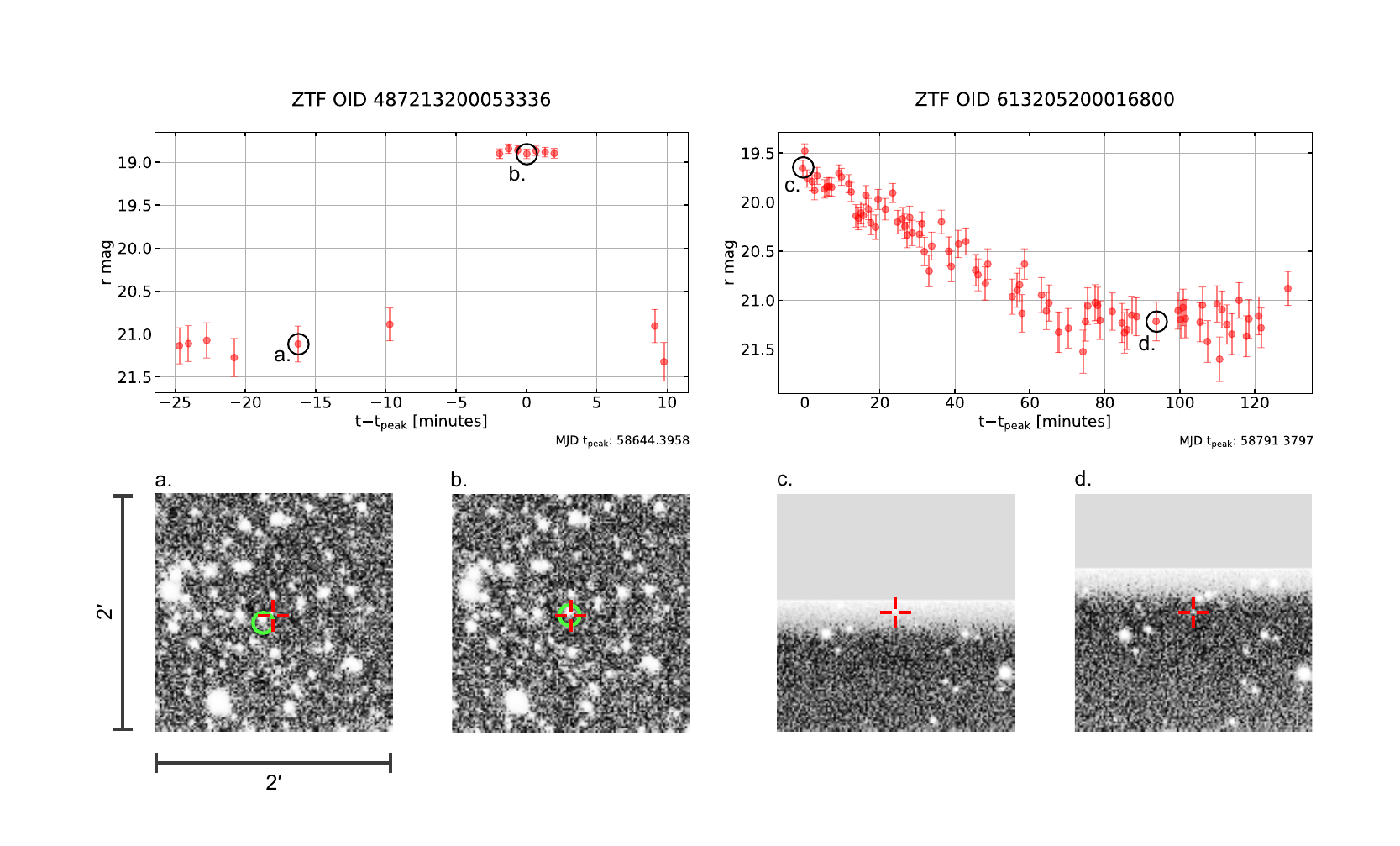}
    \caption{Light curves of a plateau (left) and an artefact (right) events discovered during the sample selection stage of flare identification. In the analysis of the light curve with a plateau in the peak, it was identified that such an increase in brightness is an overlap between the star and the asteroid 2003 XY 11. Two FITS  images with different asteroid positions are presented at \textbf{a.} and \textbf{b.} subplots. The image scale is 1$''$ to 1 px. The right-hand panel shows a light curve that visually resembles a smooth decrease typical for red-dwarf flares, but analysis of its FITS files shows that its profile is due to the frame edge being on source localization. Two fits images with different frame edge position are presented at the \textbf{c.} and \textbf{d.} subplots.}
    \label{fig:criteria_plot}
\end{figure*}

When deciding whether an event belongs to a red dwarf flare, we applied the following criteria:
\begin{itemize}
    \item The light curve profile should be  typical for a red dwarf flare: sharp increase in brightness followed by a smooth decrease. For example, the short-duration plateau observed during a ``flare'' is typically explained by the occultation of a star by an asteroid (left panel of Fig.~\ref{fig:criteria_plot}).
    \item Absence of an artefact at the frame, likesatellites, frame edges, defocusing, ghost, bright star nearby,  cosmics or hot pixels  (right panel of Fig.~\ref{fig:criteria_plot}).
    \item The object should not be a known variable star, whose observed changes in brightness are caused by other processes.
\end{itemize}

Based on these criteria, we scrutinized 343 objects of both the parametric fit and the machine learning candidates.
This resulted in a final dataset of \totcount{} flaring M-dwarf stars.
Only two flares (\texttt{762109400005614,764114400003060}) were found in ZTF $g$-band, the rest were identified in ZTF $r$-band. The remaining 209 candidates turned out to be artefacts (75 objects), instances of a star being occulted by an asteroid (128 objects) and parts of short-periodic star light-curves (6 objects).

The final sample of M-dwarf flares and their main characteristics are presented in Table~\ref{tab:Table_1}. For M-dwarfs with recurrent flares, only the flare with the largest number of  photometric measurements is listed. The first column contains the ZTF DR  object identifiers (OIDs). The equatorial coordinates ($\alpha$, $\delta$) are presented in the second and third columns, respectively. The fourth column contains the geometric distance of each object estimated with Gaia EDR3 and its uncertainty~\citep{Gaia_EDR3_dist_par}. In the fifth column, the object's $A_r$ extinction is given~\citep{2011ApJ...737..103S,Green_2019}. The sixth, seventh, eighth, and ninth columns contain peak time, full width at half maximum (FWHM), amplitude, and number of points in the flare (see Sec.~\ref{sec:flare-fit-parameters}). The tenth column presents our spectral class prediction based on Pan-STARRS photometric colours. The last column contains notes.
The next section provides detail of the analysis which led to columns four to nine.


\section{Analysis}
\label{section:analysis}

\subsection{Flare energy}

For further energy analysis, we use geometric distances derived from Gaia EDR3 \citep{Gaia_EDR3_dist_par}.
Many candidates were found to be long-distant stars, for which distances have high uncertainties.
We calculate the flare energy for the subsample of candidates whose Gaia DR3 parallax was measured with uncertainty not higher than 20\%.
We also keep candidates with enough points in the flare for higher quality of flare profile fitting.
The final subsample consists of \energycount{} flares, which were used for the energy calculation.

We assume that flare radiation could be described by an optically thick black body with a temperature of $T_\text{flare} = 9000$\,K \citep{1992ApJS...78..565H}. However, we acknowledge that this assumption is rather simplistic, as the optical continuum spectrum is significantly more complex, varying from flare to flare and from the impulsive to the gradual phases of each flare \citep{2013ApJS..207...15K,2019ApJ...871..167K,2023ApJ...944....5B}. Additionally, there is some $\textrm{H}_\alpha$ line contribution to the $r$-band during the gradual decay phase~\citep{1991ApJ...378..725H}. Nonetheless, we use this model to maintain consistency with the early assumptions made for calculating flare energies in Kepler~\citep{Shibayama_2013}, so the bolometric luminosity, $L_{\textrm {flare}}(t)$, is given by:
\begin{equation}\label{eq:l-bolometric}
    L_{\text{flare}}(t) = \rm{\sigma_{SB}} T^4_{\text{flare}} \mathcal{A}_{\text{flare}}(t)\,,
\end{equation}
where $\rm{\sigma_{SB}}$ is the Stefan--Boltzmann constant, and $\mathcal{A}_\text{flare}(t)$ is the flare surface area that changes over time.

Now our objective is to estimate $\mathcal{A}_\text{flare}(t)$ from the observed flare flux.
First, we introduce $\mathcal{A}_\perp$, a projection of $\mathcal{A}_\text{flare}$ in the picture plane.
Then, the spectral flux density of the flare $F_{\nu } = \Omega B_\nu$, where $\Omega = \mathcal{A}_\perp / d^2$ is the solid angle of the flare as observed from distance $d$, and $B_\nu$ is the black-body intensity -- Planck function.

We did not directly observe the spectral flux density with photometric surveys.
Instead, its value was averaged over the passband transmission of the photometric filter in use.
The averaged spectral flux density in ZTF $r$-band, $F_r$, is given by \citet{1986HiA.....7..833K}:

\begin{equation}
F_r =\frac{\int F_{\nu }( \nu ) / \nu \, R( \nu ) \, \mathrm{d}\nu }{\int 1/\nu \, R( \nu ) \, \mathrm{d}\nu } ~=~ \frac{\mathcal{A}_{\perp }}{d^2} \frac{\int B_{\nu } (\nu) / \nu \, R( \nu ) \, \text{d}\nu }{\int 1/\nu \, R( \nu ) \ \text{d}\nu },
\label{app_flux}
\end{equation}
where $R$ is the filter transmission function.

Using equation~\ref{app_flux} we got $\mathcal{A}_{\perp}(t)$:
\begin{equation}
\mathcal{A}_{\perp}(t) = d^2 \, \frac{F_{r}(t)}{B_{r}},
\end{equation}
where $B_r$ is the black-body intensity averaged over $r$ passband.
Being combined with  Eq.~\eqref{eq:l-bolometric} it gives the final expression of bolometric luminosity:
\begin{equation}
    L_{\text{flare}} = \sigma_\text{SB} \, T^4_{\text{flare}} \, d^2 \, \frac{\mathcal{A}_\text{flare}}{\mathcal{A}_\perp} \frac{F_{r}(t)}{B_{r}}\,.
\end{equation}
Since we do not know the shape of the flare, we introduce the geometric factor $\mathcal{A}_\text{flare} / \mathcal{A}_\perp$.
To be consistent with previous studies~\citep{Shibayama_2013,Yang2017}, we assume that this factor does not change with time and equals one.

Finally, we arrive at the expression for the bolometric energy:

\begin{equation}
    E_{\text{flare}} = \int L_{\text{flare}}(t)\,\text{d}t \,.
\end{equation}

This integral value depends only on the filter-averaged spectral flux density $F_r(t)$, which can be derived from observed magnitudes $m$. We convert observed magnitudes $m$ at time moments $t_i$ to fluxes taking into account extinction $A_r$ given by the 3-D map of Milky Way dust reddening~\citep{Green_2019}:
\begin{equation}
f(t_i) = 10^{-0.4 (m(t_i) - m_0 - A_r)}\,,
\end{equation}
where $m_0$ is the AB-magnitude zero-point.
To make the integral value more robust to observation uncertainties, we fit observations with the parametric function~\eqref{eq:mendoza} and use these fitted model fluxes as a proxy to the filter-averaged spectral flux density: $F_r(t) = f(t) - f^*$.

The energy calculation results are given inTable~\ref{table:flares_energy}.

\begin{table*}
\centering
\renewcommand{\arraystretch}{1.2}
\begin{tabular}{ccccccc}
\toprule
\midrule
\textbf{ZTF DR OID}   & \thead{\textbf{E}, \\ $10^{33}$\,erg} & \thead{\textbf{amplitude}, \\ $\Delta$ mag} & \thead{\textbf{FWHM}, \\ hours} & 
\thead{\textbf{distance}, \\ pc} & 
\thead{\textbf{spectral} \\ \textbf{class}} & \textbf{n points} \\
\midrule
\bottomrule
257209100009778 & $83.22 \pm^{14.44}_{10.53}$ & $-$3.015 & 0.23444 & $195.41^{+16.28}_{-12.78}$ & M7 & 78 \\[2px]
283211100006940 & $44.76 \pm^{29.16}_{18.37}$ & $-$2.215 & 0.04116 & $519.19^{+148.03}_{-120.53}$ & M4 & 9 \\[2px]
385209300066612 & $6.73 \pm^{2.71}_{1.55}$ & $-$2.591 & 0.04154 & $252.81^{+46.63}_{-30.98}$ & M4 & 21 \\[2px]
412207100011243 & $47.44 \pm^{13.15}_{9.05}$ & $-$3.127 & 0.07497 & $248.12^{+32.29}_{-24.91}$ & M5 & 14 \\[2px]
436207100033280 & $9.85 \pm^{2.35}_{1.77}$ & $-$2.186 & 0.06881 & $376.80^{+42.50}_{-35.43}$ & M4 & 15 \\[2px]
437212300061643 & $169.44 \pm^{18.78}_{14.95}$ & $-$4.560 & 0.10187 & $171.94^{+9.28}_{-7.76}$ & M4 & 39 \\[2px]
540208400015276 & $7.34 \pm^{2.41}_{1.61}$ & $-$2.163 & 0.04033 & $321.89^{+49.02}_{-37.51}$ & M6 & 23 \\[2px]
542214100014895 & $10.24 \pm^{0.42}_{0.43}$ & $-$2.481 & 0.08716 & $125.11^{+2.53}_{-2.64}$ & M4 & 29 \\[2px]
592208400030991 & $8.00 \pm^{0.69}_{0.60}$ & $-$3.543 & 0.05552 & $130.01^{+5.46}_{-4.98}$ & M7 & 42 \\[2px]
615214400005704 & $404.07 \pm^{195.01}_{87.72}$ & $-$2.629 & 0.26036 & $496.64^{+108.08}_{-57.20}$ & M4 & 20 \\[2px]
726209400028833 & $27.66 \pm^{0.82}_{0.80}$ & $-$1.782 & 0.12602 & $162.36^{+2.38}_{-2.37}$ & M4 & 65 \\[2px]
768211400063696 & $4.45 \pm^{1.30}_{1.23}$ & $-$2.043 & 0.07152 & $243.37^{+33.22}_{-36.45}$ & M6 & 18 \\[2px]
771216100033044 & $152.34 \pm^{27.38}_{31.06}$ & $-$2.105 & 0.18924 & $450.52^{+38.82}_{-48.54}$ & M3 & 33 \\[2px]
804211400018421 & $36.73 \pm^{20.97}_{9.65}$ & $-$1.944 & 0.15684 & $495.73^{+125.58}_{-70.05}$ & M4 & 19 \\
\bottomrule
\bottomrule
\end{tabular}
\caption{Bolometric energy estimations for the subsample of \energycount{} flare candidates. The upper and lower energy errors were defined according to the Gaia EDR3 estimated uncertainties on geometric distances. The values of estimated amplitude, FWHM, number of points in the flare (described in details in Sec.~\ref{sec:flare-fit-parameters}), Gaia EDR3 geometrical distance and most probable spectral class (see Sec.~\ref{sec:spec-class}) are also specified.
}
\label{table:flares_energy}
\end{table*}

\subsection{Flare fit parameters}
\label{sec:flare-fit-parameters}

We use the same parametric function from \eqref{eq:mendoza} to estimate the amplitude, FWHM and number of points in the flare (Table~\ref{tab:Table_1}). Firstly, a manual operation of flares' observations subtraction from the full light curve was conducted -- each stars' light curve was trimmed to capture the flare itself and some amount of quiescent points which are necessary for the further fitting operation.

Secondly, for each flare, its profile was fitted using a parametric function (Equation \eqref{eq:mendoza}) to get a continuous representation convenient for further analysis.
For complex light curves with several flare peaks, only the peak with the maximum amplitude (main peak) is considered for further fitting.
The amplitude is calculated as the difference between the maximal and minimal model flux, which is achieved over a time interval.
As a next step, the FWHM is measured as the difference between time points where the model curves possess values equal to half of the defined amplitude.

In order to define the number of points in the flare, we adopted the following criterion: all points of the light curve whose observed flux exceeds $f^*$ (quiescence stellar flux obtained from the model) by $3\sigma$, where $\sigma$ is the mean observational error of the selected part of the light curve, were considered as belonging to the flare.

\subsection{Spectral class}
\label{sec:spec-class}

Due to the lack of spectral information for most of our objects, we employ a photometric approach to estimate their spectral types. We use stacked magnitudes from Pan-STARRS DR2 \citep{2016Pan-starrs1,2018AAS...23143601F} to build the ($r-i$, $i-z$) colour-colour diagram of our sample.
The method is adopted from \cite{Kowalski_2009} and allowed us to define the spectral subtypes of M-dwarfs based only on their photometric colours. 

First, we correct the colours for the galactic reddening.
We use extinction values and coefficients from two different sources: the three-dimensional map of Milky Way dust reddening ``Bayestar19'' \citep{Green_2019}, and, if this map does not contain the object, the map of Galactic Dust Reddening and Extinction presented by \citealt{2011ApJ...737..103S}. Based on these maps, we calculate the final extinction value using:
\begin{equation}
    \boldsymbol{A} = E \boldsymbol{R}
\end{equation}
where $\boldsymbol{A} = (A_r, A_i, A_z)$ represents the final extinction value in each filter, $\boldsymbol{R} = (R_r, R_i, R_z)$ is the “extinction vector”, relating a scalar reddening to the extinction in each passband and $E$ is the reddening in the dust-map specific units.
The extinction vector is $\boldsymbol{R} = (2.617, 1.971, 1.549)$ for the 3-D map ``Bayestar19'', and $\boldsymbol{R} = (2.271, 1.682, 1.322)$ for the 2-D map of Galactic Dust Reddening and Extinction.
For the 3-D map, we use Gaia EDR3 geometrical distance as discussed above.

\cite{Kowalski_2009} proposed a table with the best-fit parameters of the two-dimensional Gaussian distribution for each M-dwarf spectral subtype in the ($r-i$, $i-z$) colour space.
We calculate the probability of belonging to the corresponding spectral subclass following their proposal, 
\begin{equation}
 p_m(x) = \frac{1}{2\pi \Vert \Sigma_m \Vert^{1/2}}\exp(-\frac{1}{2}(\boldsymbol{x}-\boldsymbol{\mu}_m)^T\Sigma_m^{-1}(\boldsymbol{x}-\boldsymbol{\mu}_m))\,,
\end{equation}
where $m$ is an M-dwarf spectral subclass index, from 0 to 7, $\boldsymbol{\mu}_m$ and $\Sigma_m$ are Gaussian mean values and covariance matrix for this subclass, $\boldsymbol{x} = ({r-i}, {i-z})$ are deredded colors of the studied star.

\begin{figure}
\center{\includegraphics[width=90mm,scale=0.8]{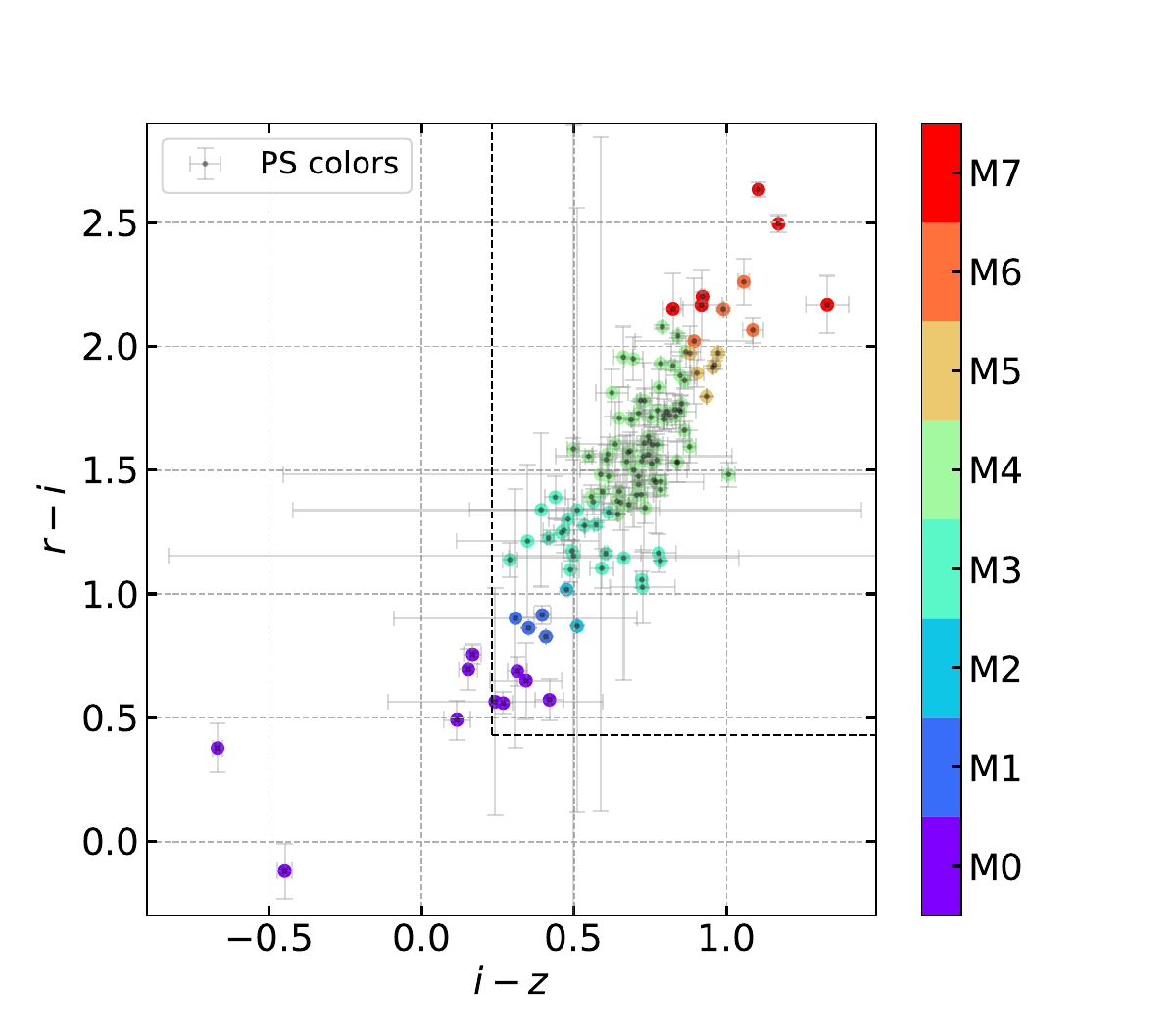}}
\caption{Spectral classes distribution calculated based on the photometric data from Pan-STARRS DR2. The black dashed lines show the blue limits of color indexes for the M-dwarf spectral type according to~\citealt{Kowalski_2009}. Five points that fall outside these limits likely belong to a different spectral class.
}
\label{fig:Spectral_Classes}
\end{figure}

For each star, we assign the spectral subclass $m$ corresponding to the maximum probability $p_m$.
We show the objects in the colour-colour diagram (Figure~\ref{fig:Spectral_Classes})  with subclass corresponding to the point estimations of the pair of object's colours.
However, the deredded colour values may have large uncertainties associated with Pan-STARRS photometry error, distance estimation error (applicable only for the 3-D dust maps), and extinction errors.

Finally, we apply a colour cut based on \cite{West2000} and \cite{Bochanski_2007}, to construct a more pure M-dwarfs sample. \cite{Kowalski_2009} presented the following  limits, which take extinction into account: $(r-i)  > 0.42$,  $(i-z)> 0.23$.

Starting from the \totcount{} stars found, we determine the spectral subclasses of $132$ of them (see Table~\ref{tab:Table_1}, Figure~\ref{fig:Spectral_Classes}). Five objects that lie outside of the M-dwarf limits might have a different spectral type. Their OIDs are \texttt{435211200068171}, \texttt{540215200069194}, \texttt{687207100049742}, \texttt{687214100050598}, and \texttt{768209200100383}, with corresponding $A_r$ values of 1.14, 5.11, 1.15, 2.82, 2.67, respectively. It is possible that the interstellar reddening for these objects has been overestimated: four of them do not have Gaia distances, and for another, the distance is measured with high uncertainty, which may cause the object to appear shifted towards bluer colors.

For the 16 brightest objects, Gaia DR3 provides effective temperature estimations~\citep{2023A&A...674A...1G}.
We use these estimations to define a spectral class based on the relationship between the stellar effective temperature and the spectral class for main-sequence stars adopted from~\cite{Malkov_2020}.
In all cases, this method confirms the spectral class M for these objects.
For seven objects, we have an exact agreement for the spectral subclass between this method and the method based on Pan-STARRS colours.
There is also one object (\texttt{719206100008051}) with available spectrum from SDSS DR18~\citep{SDSS_DR18} data and the corresponding spectral classification, which is consistent with our colour-based classification.
See the individual object information in the ``note'' column of Table~\ref{tab:Table_1}.

It should be noted that using Gaia geometric distances, we can estimate the absolute magnitude of our objects. The objects having much brighter absolute magnitude than expected for main-sequence M-dwarf stars~\citep{main_sequence} also have large parallax uncertainties. We mark these objects with crosses in the distance column of Table~\ref{tab:Table_1}. Although most of those with more confident distance estimations are consistent with the main sequence, the outliers may be explained by system multiplicity or other systematic errors.


\section{Discussion}
\label{section:discussion}

\subsection{Flares morphology}

The flaring stars we study in this paper vary in the number of observational points per flare, flare recurrence, and light-curve profiles. The light curves of all the flares are provided in the supplementary material.

\begin{figure*}
\centering
	\includegraphics[width=\textwidth]{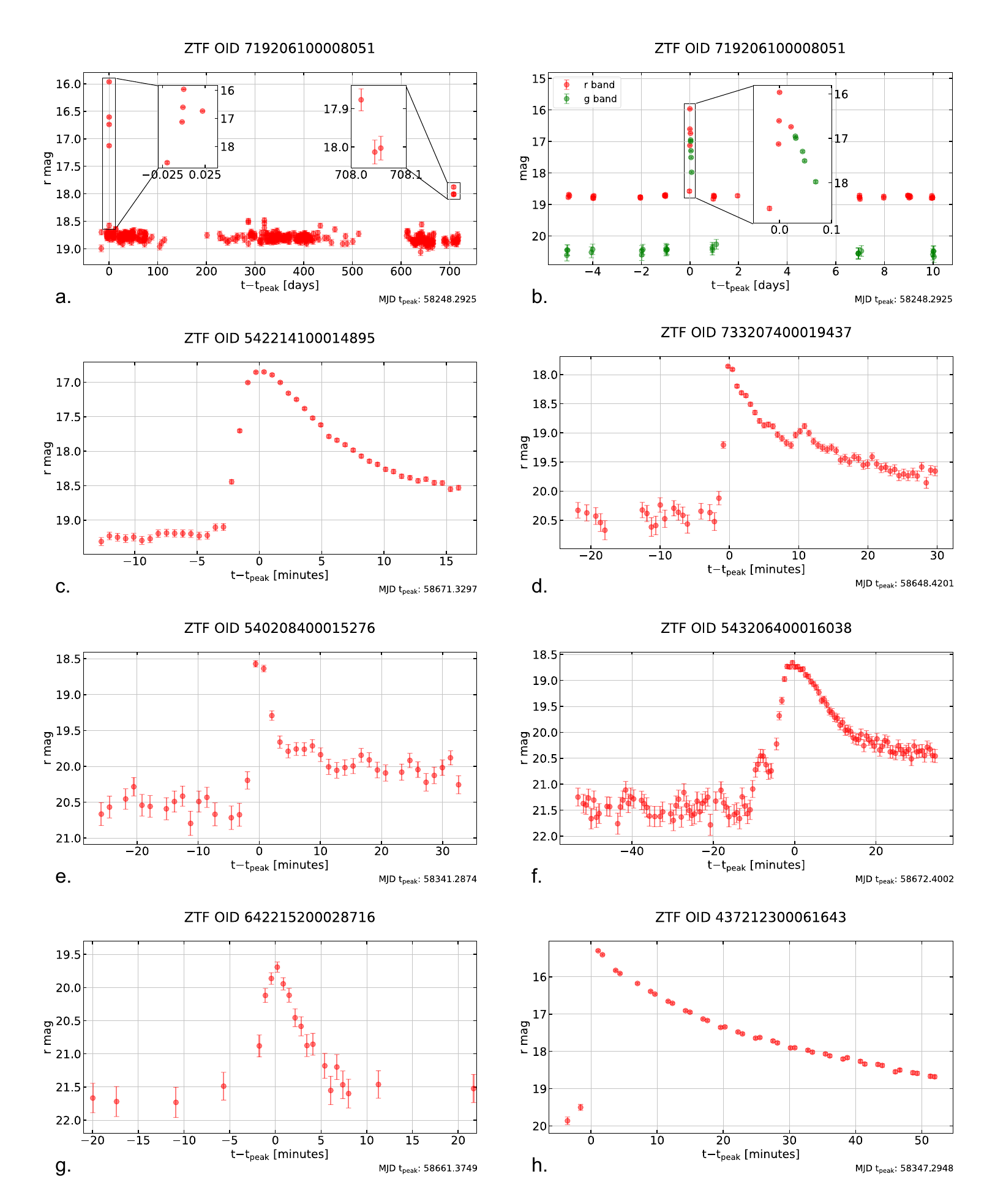}
    \caption{Light curves of M-dwarf flares with different morphology discovered with AAD (\textbf{a}-\textbf{b}, \textbf{h}) and parametric fit methods (\textbf{c}-\textbf{g}). \textbf{a.} Recurrent flares with zoomed-in flares in the inset plots. \textbf{b.} Simultaneous flares in $g$- and $r$-bands with zoomed-in flares in the inset plot.
    \textbf{c.} Classical singular flare with rapid rise and smooth decline. \textbf{d.} Complex flare with two components. \textbf{e.} Complex flare with multiple components. \textbf{f.} Complex flare with variable background, where a less energetic peak precedes the most energetic one. \textbf{g.} Flare with symmetric light-curve profile. \textbf{h.} Flare with the highest magnitude amplitude.}
    \label{fig:all_plots}
\end{figure*}

For some of the flaring M-dwarfs multiple flares are observed: either multiple flares in a single passband (Fig.~\ref{fig:all_plots}a), or a single flare simultaneously observed in $g$- and $r$-bands (Fig.~\ref{fig:all_plots}b).  

Additionally, we distinguish between classical and complex flare events on the basis of their temporal structure (e.g.,~\citealt{2013ApJS..207...15K,2014ApJ...797..122D}). Classical flares have a single-peak profile, characterized by fast rise and exponential decay (Fig.~\ref{fig:all_plots}c). However, the majority of our flares display a much more complex structure. This complexity ranges from relatively simple flares (see Fig.~\ref{fig:all_plots}d) to highly complex flares consisting of multiple components (see Fig.~\ref{fig:all_plots}e). According to \citet{2016IAUS..320..128D}, studying such flare complexity could clarify their origins, as it remains uncertain whether they are produced by a single active region or by triggering separate nearby regions. In the first scenario, \citet{2016IAUS..320..128D} anticipates that the most energetic flare occurs first, followed by a sequence of less energetic events. However, this is not supported by some flares in our sample, where a less energetic peak precedes the most energetic one (Fig.~\ref{fig:all_plots}f).

Despite our best efforts, we recognize  that there is still some inherited uncertainty in the classifications presented here, especially when only one point is available. They could be associated with hot spots on a stellar surface, self-lensing binaries or other types of stars that flare (e.g.,~\citealt{2024arXiv240207885K}). As it was recently stated by \citealt{2023arXiv231117862C}, a possible scenario would be self-lensing detached binaries, containing a stellar-mass neutron star or a black hole. The brightening occurs when the compact object transits the companion star. In that case, a symmetric light curve should be observed. Among our flares, there are a few candidates which  satisfy this criteria (e.g., Fig.~\ref{fig:all_plots}g).

\subsection{Parametric fit vs AAD}

Searching for flares on M-dwarfs is a complex task. Each data set, whether it comes from different surveys or not, is unique. That is why it is so important to explore various searching strategies. In this paper, we apply two different methods for M-dwarf flares search. Below we compare both approaches. 

\begin{figure}
     \centering
     \begin{subfigure}{\columnwidth}
         \centering
         \includegraphics[width=\textwidth]{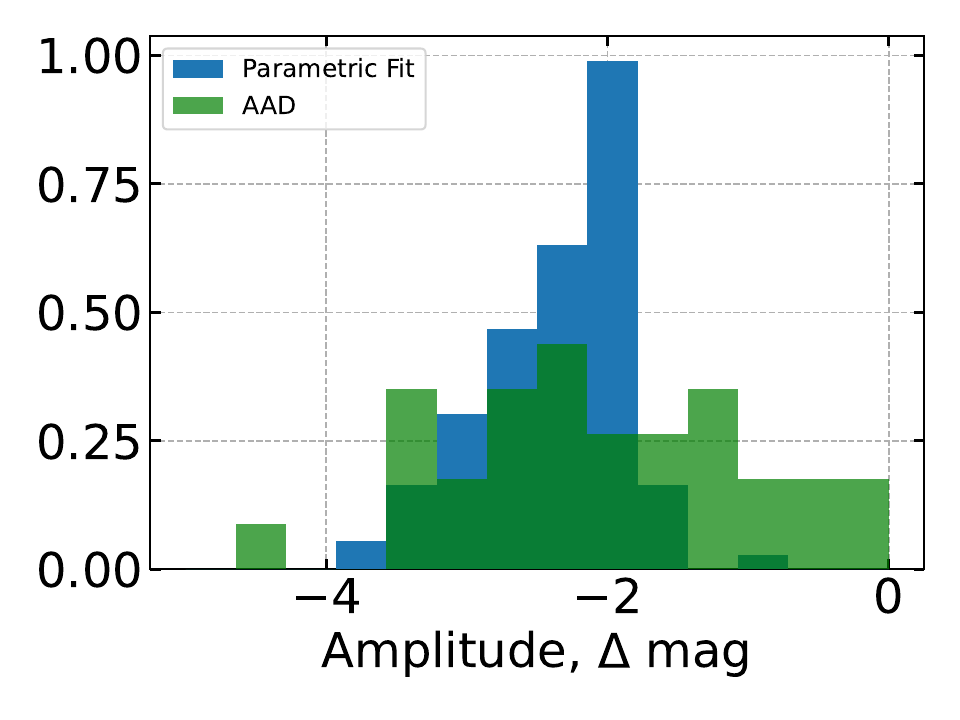}
         \caption{}
         \label{fig:ampl_hist}
     \end{subfigure}
     \hfill
     \begin{subfigure}{\columnwidth}
         \centering
         \includegraphics[width=\textwidth]{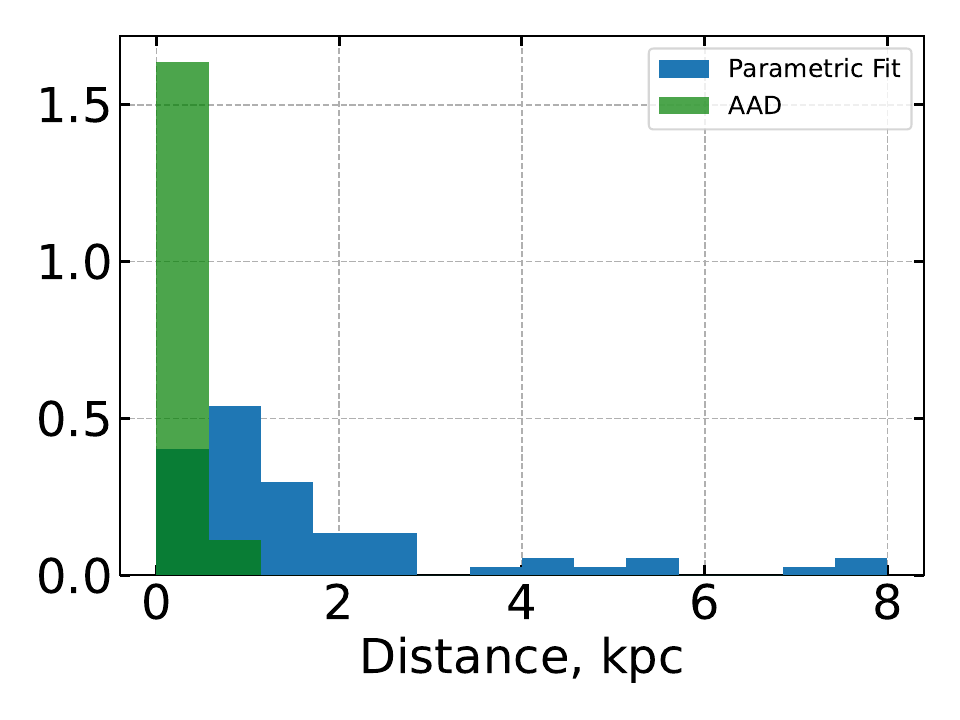}
         \caption{}
         \label{fig:dist_hist}
     \end{subfigure}
    \caption{(a) Normalized histogram of amplitudes of flares found with parametric fit and AAD methods. (b) Normalized histogram of distances to M dwarfs found with parametric fit and AAD methods.}
\end{figure}

First of all, the parameters associated with the discovered flares are different. Due to the ZTF's sporadic observation schedule, flares found with AAD are sampled from incomplete light curves, often missing the peak brightness. In contrast, a parametric fit was applied to high-cadence data, thereby providing a well-defined light-curve profile. Consequently, flares found with AAD algorithm have systematically smaller amplitudes in comparison with the ones from the parametric fit (see Fig.~\ref{fig:ampl_hist}). The number of points in AAD flares is significantly lower, which makes them less reliable and more difficult to confirm. Also, recurrent flares were more often found by the AAD approach, since it had access to long-duration light curves (e.g., Fig.~\ref{fig:all_plots}a).

\begin{figure*}
\centering
	\includegraphics[width=170mm,scale=0.8]{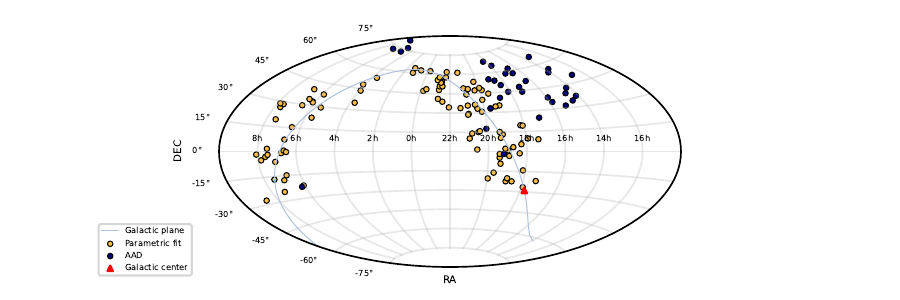}
    \caption{Spatial distribution of M dwarfs with flares in equatorial coordinates. Dark blue points indicate the flares found with AAD method, the yellow ones are flares detected with parametric fit search. Thin blue line shows the Galactic plane and red triangle is center of the Galaxy.}
    \label{fig:map}
\end{figure*}

Secondly, regarding the spatial distribution of flares, the parametric fit search was limited by a ZTF high-cadence coordinate cut, resulting in flares that are located within the Milky Way plane. In contrast, the AAD does not have any coordinate restriction, yet the observed bias towards higher galactic latitudes in the Fig.~\ref{fig:map} can be attributed to the selection effect, i.~e. a fewer runs of the algorithm in fields within or close to the galactic plane, since this method requires expert evaluation at every step of its operation. Moreover, flares obtained by the parametric fit method have a larger distance to Earth according to Gaia EDR3 (Fig.~\ref{fig:dist_hist}). This can be explained by the use of a  reduced $\chi^2$ metric to evaluate goodness of fit, which includes photometric errors, therefore lower metric quantities are systematically associated with fainter and more distant sources.

Finally, since AAD is an anomaly detection algorithm, it was not originally intended for flare detection. However, by treating flares as anomalies, we are able to successfully adapt it to search for flares. It is interesting to note that only one over a few high-cadence flares identified by AAD exhibit exceptionally high amplitudes when compared to typical flare energy outputs (Fig.~\ref{fig:all_plots}, see also \citealt{2020ApJ...892..144R,2022Ge&Ae..62..911G}). On the other hand, the parametric fit technique excels in isolating well-sampled flares within high-cadence amplitudes, enabling the observation of complex substructures within the light curves (e.g, Fig.~\ref{fig:all_plots}e). It allows performing a comprehensive analysis of individual flares~\citep{2016IAUS..320..128D}.


\section{Conclusions}

This paper is devoted to the search for M-dwarf flares in the eighth-data release of the Zwicky Transient Facility survey. We explore two different approaches: a parametric fit search and a machine learning method.

We visually scrutinized \scrutinisedcount{} candidates, filtering out artefacts and known variables of other types, to identify \totcount{} M-dwarf flares. This constitutes the largest sample of ZTF M-dwarf flares identified to date. The associations with the M spectral class are confirmed through the ($r-i$, $i-z$) colour diagram analysis, though some classifications may be incorrect, and we note  opportunities for identifying exotic events like self-lensing binaries. For \energycount{} objects, we calculate the flare energy, ranging from $\sim$7$\times10^{33}$ to $\sim$404$\times10^{33}$ erg, which is consistent with the higher end of the energy distribution reported in the literature~\citep{Yang2017}.

The comparison between the two approaches shows that each of them identifies flares of different parameters and distribution in the sky. For example, the parametric fit search found fainter flares, while AAD, despite lagging in flare light curve quality, identified recurrent flares. Additionally, the highest amplitude flare in the sample was discovered using AAD. Since each method has its own limitations,  diverse strategies for flare detection are necessary to form a comprehensive picture of these phenomena.

Although the ZTF survey is not specifically designed for fast transients due to its 2-3 day cadence, it conducts private high-cadence observational campaigns. Such campaigns are also envisaged by the observational strategy of the Vera Rubin Observatory Legacy Survey of Space and Time with Deep Drilling Fields program \cite{lsst}. For the search and study of red dwarfs, we should not rely solely on dedicated surveys; instead, we must learn to extract necessary information from surveys not originally intended for this purpose. Therefore, developing methods for data filtering and flare identification is highly relevant.



\section*{Acknowledgements}
A.~Lavrukhina, M.~Kornilov, A.~Volnova and T.~Semenikhin acknowledges support from a Russian Science Foundation grant 24-22-00233, https://rscf.ru/en/project/24-22-00233/. Support was provided by Schmidt Sciences, LLC. for K.~Malanchev. V.~Krushinsky acknowledges support from the youth scientific laboratory project, topic FEUZ-2020-0038.

This work has made use of data from the European Space Agency (ESA) mission {\it Gaia} (\url{https://www.cosmos.esa.int/gaia}), processed by the {\it Gaia} Data Processing and Analysis Consortium (DPAC, \url{https://www.cosmos.esa.int/web/gaia/dpac/consortium}). Funding for the DPAC has been provided by national institutions, in particular the institutions participating in the {\it Gaia} Multilateral Agreement. 

The Pan-STARRS1 Surveys (PS1) and the PS1 public science archive have been made possible through contributions by the Institute for Astronomy, the University of Hawaii, the Pan-STARRS Project Office, the Max-Planck Society and its participating institutes, the Max Planck Institute for Astronomy, Heidelberg and the Max Planck Institute for Extraterrestrial Physics, Garching, The Johns Hopkins University, Durham University, the University of Edinburgh, the Queen's University Belfast, the Harvard-Smithsonian Center for Astrophysics, the Las Cumbres Observatory Global Telescope Network Incorporated, the National Central University of Taiwan, the Space Telescope Science Institute, the National Aeronautics and Space Administration under Grant No. NNX08AR22G issued through the Planetary Science Division of the NASA Science Mission Directorate, the National Science Foundation Grant No. AST-1238877, the University of Maryland, Eotvos Lorand University (ELTE), the Los Alamos National Laboratory, and the Gordon and Betty Moore Foundation.

This work has made use of data from ZTF, supported by the National Science Foundation under Grants No. AST-1440341 and AST-2034437 and a collaboration including current partners Caltech, IPAC, the Oskar Klein Center at Stockholm University, the University of Maryland, University of California, Berkeley , the University of Wisconsin at Milwaukee, University of Warwick, Ruhr University, Cornell University, Northwestern University and Drexel University. Operations are conducted by COO, IPAC, and UW.

\section*{Data Availability}

The ZTF light-curve data underlying this article are available in the NASA/IPAC Infrared Science Archive\footnote{\url{https://irsa.ipac.caltech.edu/}}. The final table with selected flares and their parameters are given in Appendix~\ref{section:catalogue}.


\bibliographystyle{mnras}
\bibliography{rdf}


\appendix

\onecolumn

\section{Table of M-dwarf flares}\label{section:catalogue}
We show here the table of all found flares and the corresponding stars with their main characteristics.
\renewcommand{\arraystretch}{1.39}
\begin{landscape}
\begin{longtable}{cccccccccc|p{40mm}}
\toprule
\toprule
\textbf{ZTF DR OID} & \thead{$\bm{\alpha}$, \\ deg} & \thead{$\bm{\delta}$, \\ deg} & \thead{\textbf{distance}, \\ pc} &\thead{$\bm{A_r}$$^1$, \\ (mag)} & \thead{$\bm{t}_\mathrm{\bf peak}$$^2$, \\ MJD$-$58000} & \thead{\textbf{FWHM}$^2$, \\ hours} & \thead{\textbf{amplitude}$^{2}$, \\ $\Delta$ mag} & \textbf{n points}$^2$ & \thead{\textbf{spectral} \\ \textbf{class}} & \thead{\textbf{note}} \\
\midrule
\multicolumn{10}{c}{\textbf{AAD method}}\\
\bottomrule
257209100009778 & 92.9219 & $-$22.7911 & $195.41^{+16.28}_{-12.78}$ & 0.0000 & 471.3549 & 0.23444 & $-$3.015 & 78 & M7 &  \\
437212300061643 & 287.9685 & $-$1.9057 & $171.94^{+9.28}_{-7.76}$ & $0.0109^{+0.1871}_{-0.0109}$ & 347.2950 & 0.10187 & $-$4.560 & 39 & M4 & Effective temperature available $2955.8^{+10.6}_{-19.6}$ K, M5 \\
592208400030991 & 300.7593 & 17.5055 & $130.01^{+5.46}_{-4.98}$ & 0.0000 & 343.2278 & 0.05552 & $-$3.543 & 42 & M7 &  \\
634207400007102 & 255.6085 & 24.6610 & $590.71^{+109.63}_{-89.24}$ & 0.1832 & 219.4415 & -- & $-$3.456 & 6 & M4 &  \\
676211100006667 & 218.9423 & 34.4342 & $44.91^{+0.23}_{-0.19}$ & 0.0000 & 217.3884 & -- & $-$0.910 & 5 & M3 &  \\
677206300030165 & 228.0519 & 31.6719 & $94.94^{+1.53}_{-1.60}$ & 0.0000 & 217.3703 & -- & $-$3.097 & 9 & M6 & Recurrent flares in $g$- and $r$- bands \\
678210100002177 & 237.5196 & 34.9555 & $275.84^{+2.79}_{-3.29}$ & 0.1308 & 350.1435 & -- & $-$0.620 & 4 & M3 & Recurrent flares in $g$- and $r$- bands, effective temperature available $3408.1^{+3.7}_{-3.7}$ K, M2 \\
718201300005383 & 212.5848 & 37.0767 & $76.52^{+0.29}_{-0.26}$ & 0.0000 & 226.2941 & -- & $-$0.982 & 5 & M4 & Recurrent flares in $g$- and $r$- bands, simultaneous flares in both bands \\
719206100008051 & 219.6046 & 39.6384 & $126.23^{+6.16}_{-4.84}$ & 0.0000 & 248.2924 & -- & $-$2.602 & 4 & M5 & Recurrent flares in $g$- and $r$- bands, simultaneous flares in both bands, spectrum available,  SDSS J143825.07+393819.5, M5 \\
719216300003437 & 213.9701 & 42.9291 & $391.43^{+4.99}_{-4.74}$ & 0.0000 & 222.3259 & -- & $-$0.222 & 4 & M2 & Flares in $g$- and $r$- bands, effective temperature available $3592.5^{+11.1}_{-2.1}$ K, M1 \\
721201200001366 & 238.6182 & 38.3407 & $101.37^{+1.47}_{-1.50}$ & 0.0000 & 216.3633 & -- & $-$2.099 & 3 & M5 &  \\
726209400028833 & 282.5514 & 40.8524 & $162.36^{+2.38}_{-2.37}$ & 0.0000 & 324.3501 & 0.12602 & $-$1.782 & 65 & M4 & Effective temperature available $3231.9^{+5.5}_{-5.3}$ K, M3 \\
756211200000623 & 192.6775 & 49.4891 & $192.50^{+10.25}_{-9.49}$ & 0.0000 & 217.2386 & -- & $-$2.344 & 1 & M4 & Flares in $g$- and $r$- bands \\
761214100001680 & 245.8604 & 51.3084 & $274.63^{+12.79}_{-10.89}$ & 0.0000 & 216.3775 & -- & $-$2.765 & 5 & M4 & Effective temperature available $3093.7^{+3.1}_{-1.9}$ K, M4 \\
762109400005614 & 257.9108 & 47.7715 & $126.54^{+0.46}_{-0.43}$ & 0.0000 & 635.4311 & -- & $-$1.379 & 4 & M3 &  \\
762201400007313 & 258.2507 & 44.1310 & $150.43^{+2.80}_{-2.91}$ & 0.0000 & 377.1626 & -- & $-$3.470 & 3 & M4 & Recurrent flares in $r$-band, one simultaneous flare in both bands, effective temperature available $2966.4+^{9.7}_{-6.8}$ K, M5 \\
764114400003060 & 275.3463 & 50.2501 & $238.04^{+1.09}_{-1.22}$ & 0.0000 & 691.3068 & -- & $-$0.276 & 1 & M1 & Effective temperature available $3805.7^{+6.6}_{-4.5}$ K, M0 \\
764203100012551 & 271.8373 & 44.9095 & $480.39^{+46.77}_{-28.06}$ & $0.1308_{-0.0205}$ & 198.5153 & -- & $-$1.977 & 3 & M2 & Effective temperature available $3249.5^{+35.6}_{-22.3}$ K, M3 \\
791209200005999 & 205.9704 & 55.9733 & $289.52^{+10.94}_{-13.54}$ & 0.0000 & 248.2657 & -- & $-$1.468 & 4 & M4 & Recurrent flares in $r$-band, effective temperature available $3210.1^{+8.7}_{-8.8}$ K, M4 \\
792207200006505 & 211.3704 & 54.1319 & $93.68^{+0.28}_{-0.34}$ & 0.0000 & 258.2268 & -- & $-$1.526 & 4 & M4 & Recurrent flares in $g$- and $r$- bands, one simultaneous flare in both bands, effective temperature available $3236.1^{+1.9}_{-4.0}$ K, M3 \\
795213200016815 & 251.0949 & 57.8094 & $153.76^{+2.51}_{-1.90}$ & $0.0124^{+0.0074}_{-0.0057}$ & 262.3504 & -- & $-$2.332 & 2 & M4 & Recurrent flares in $g$- and $r$- bands, effective temperature available $3169.7^{+4.5}_{-3.9}$ K, M4 \\
796214100003950 & 259.9339 & 58.2575 & $33.87^{+0.03}_{-0.04}$ & 0.0000 & 379.2584 & -- & $-$1.080 & 1 & M5 & Recurrent flares in $g$- and $r$- bands, effective temperature available $2974.1^{+4.2}_{-7.6}$ K, M5 \\
798207400001244 & 279.1108 & 53.9023 & -- & 0.0899 & 318.3486 & -- & $-$0.530 & 4 & M1 & Recurrent flares in $g$- and $r$- bands, simultaneous flares in both bands \\
798209400009221 & 284.9472 & 55.2334 & $264.90^{+18.62}_{-13.75}$ & 0.0000 & 198.5135 & -- & $-$2.167 & 3 & M4 & Flares in both $g$- and $r$- bands \\
821216100003336 & 200.1288 & 65.2912 & $282.43^{+5.75}_{-5.59}$ & 0.0000 & 353.1419 & -- & $-$1.393 & 2 & M4 & Effective temperature available $3285.1^{+21.8}_{-7.5}$ K, M3 \\
824205200007029 & 250.7534 & 61.4259 & $397.62^{+60.71}_{-48.00}$ & 0.1570 & 377.1581 & -- & $-$2.330 & 1 & M4 &  \\
825213100013108 & 267.8115 & 64.9554 & $186.73^{+10.76}_{-8.28}$ & 0.0000 & 325.2432 & -- & $-$3.250 & 4 & M4 &  \\
848205100005466 & 274.1424 & 68.5419 & $91.45^{+0.41}_{-0.53}$ & 0.0000 & 385.1791 & -- & $-$2.737 & 3 & M4 & Recurrent flares in $g$-band and a flare in $r$-band \\
857207100012456 & 81.2459 & 76.2029 & $242.02^{+3.42}_{-4.33}$ & $0.2094_{-0.0429}$ & 358.4232 & -- & $-$2.367 & 3 & M3 & Effective temperature available $3364.6^{+4.4}_{-5.1}$ K, M3 \\
858204400004738 & 100.5724 & 73.0318 & $398.96^{+27.84}_{-18.38}$ & 0.2617 & 229.1904 & -- & $-$1.826 & 1 & M2 & Recurrent flares in $r$-band, effective temperature available $3491.1^{+12.5}_{-12.4}$ K, M2 \\
858213100000788 & 126.4006 & 79.6325 & $638.21^{+61.31}_{-53.07}$ & 0.1308 & 464.3079 & -- & $-$1.194 & 1 & M3 & Effective temperature available $3392.0^{+23.8}_{-25.3}$ K, M2 \\
1866210400023756 & 78.2604 & 73.1147 & $59.19^{+0.42}_{-0.43}$ & 0.0000 & 774.3110 & -- & $-$2.545 & 3 & M7 & Recurrent flares in $r$-band \\
\midrule
\multicolumn{10}{c}{\textbf{Parametric fit method}}\\
\midrule
257214400014856 & 91.0585 & $-$22.0385 & $^\dagger 1392.66^{+1373.00}_{-707.61}$ & 0.1308 & 468.3190 & 0.09414 & $-$1.630 & 19 & M4 &  \\
260208100017563 & 109.4550 & $-$24.6727 & $^\dagger 1237.22^{+596.11}_{-397.29}$ & $0.2355^{+0.0662}_{-0.0785}$ & 493.3462 & 0.04697 & $-$2.540 & 32 & M4 &  \\
262201300031816 & 129.7493 & $-$27.8083 & -- & 0.2542 & 493.3885 & 0.15505 & $-$2.720 & 33 & M7 &  \\
280214400089687 & 259.6615 & $-$21.6633 & $^\dagger 702.36^{+293.14}_{-188.09}$ & $1.3500^{+0.1417}_{-0.1083}$ & 303.2693 & 0.04313 & $-$1.994 & 30 & M4 &  \\
281201100016537 & 268.6979 & $-$26.5830 & $^\dagger 1223.03^{+1021.79}_{-669.60}$ & $2.3154^{+0.2766}_{-1.9040}$ & 636.3873 & 0.24195 & $-$2.056 & 36 & M3 &  \\
283211100006940 & 279.3314 & $-$22.5463 & $519.19^{+148.03}_{-120.53}$ & $0.4972^{+0.0523}_{-0.0262}$ & 312.2717 & 0.04116 & $-$2.215 & 9 & M4 &  \\
284212100096997 & 284.2108 & $-$22.6351 & $^\dagger 5038.87^{+2562.02}_{-2095.08}$ & 0.6804 & 287.3557 & 0.07169 & $-$2.316 & 24 & M4 &  \\
308214300027206 & 100.8986 & $-$14.8462 & $^\dagger 707.18^{+331.81}_{-192.67}$ & $0.3921^{+0.3930}_{-0.1304}$ & 464.3919 & 0.03514 & $-$2.554 & 34 & M4 &  \\
309208200034266 & 104.3670 & $-$17.7464 & -- & 1.5852 & 464.3837 & 0.09362 & $-$2.479 & 13 & M0 &  \\
310212300021722 & 111.9763 & $-$16.8197 & $^\dagger 1145.06^{+547.17}_{-419.21}$ & $0.5234^{+0.5234}_{-0.0262}$ & 475.4510 & 0.05547 & $-$2.474 & 37 & M3 &  \\
332213200128168 & 271.9229 & $-$14.1020 & $^\dagger 1548.68^{+2089.12}_{-852.08}$ & $1.3555^{+2.3197}_{-0.2302}$ & 637.3692 & 0.29056 & $-$1.916 & 123 & M2 &  \\
334203400074200 & 283.5040 & $-$20.3306 & $^\dagger 5221.32^{+2086.45}_{-2273.88}$ & $0.8374_{-0.1832}$ & 320.3401 & 0.03386 & $-$2.877 & 17 & M4 &  \\
336202400036948 & 299.2203 & $-$20.7151 & $^\dagger 375.64^{+109.16}_{-75.84}$ & $0.4972_{-0.0766}$ & 667.4013 & 0.06337 & $-$2.468 & 41 & M4 &  \\
336212400006103 & 295.0432 & $-$16.2609 & -- & 0.3741 & 667.3851 & -- & $-$0.827 & 16 & M4 &  \\
367206100004253 & 161.5502 & $-$10.4167 & $174.61^{+12.73}_{-10.62}$ & 0.0000 & 511.2607 & 0.04158 & $-$2.129 & 18 & M7 &  \\
385209300066612 & 290.2818 & $-$9.3424 & $252.81^{+46.63}_{-30.98}$ & $0.4304^{+0.0930}_{-0.2074}$ & 292.4055 & 0.04154 & $-$2.591 & 21 & M4 &  \\
410210100030608 & 101.6796 & $-$1.0568 & -- & 2.3353 & 812.4599 & 0.06651 & $-$2.988 & 33 & M0 &  \\
410215100032143 & 99.6352 & 0.5756 & $^\dagger 1214.56^{+1149.74}_{-819.13}$ & $0.5856^{+0.5397}_{-0.5594}$ & 812.4650 & 0.03854 & $-$2.172 & 20 & M4 &  \\
410216400016069 & 97.7781 & $-$0.3300 & $^\dagger 2845.78^{+2208.58}_{-1154.52}$ & $1.6008^{+0.4928}_{-0.3084}$ & 812.4972 & 0.08735 & $-$1.797 & 24 & M3 &  \\
411203400031073 & 106.8275 & $-$6.3215 & $^\dagger 4538.29^{+3644.30}_{-2657.60}$ & $1.2896^{+0.3330}_{-1.0279}$ & 457.3170 & 0.05976 & $-$2.394 & 36 & M4 &  \\
412201100010804 & 117.6872 & $-$5.1527 & -- & 0.2787 & 456.4139 & 0.21868 & $-$2.232 & 41 & M2 &  \\
412207100011243 & 114.1311 & $-$3.1612 & $248.12^{+32.29}_{-24.91}$ & $0.0206^{+0.0056}_{-0.0206}$ & 456.5040 & 0.07497 & $-$3.127 & 14 & M5 &  \\
412212400027889 & 112.2499 & $-$1.9847 & -- & 0.2395 & 457.4206 & 0.05865 & $-$3.826 & 21 & M4 &  \\
413211400001358 & 121.0392 & $-$1.7560 & $^\dagger 1436.95^{+895.33}_{-590.04}$ & 0.0785 & 775.4796 & 0.10116 & $-$1.873 & 65 & M4 &  \\
435211200068171 & 275.6615 & $-$1.4027 & -- & 2.5958 & 640.4373 & -- & $-$1.805 & 1 & M0 &  \\
436207100033280 & 283.7603 & $-$3.3619 & $376.80^{+42.50}_{-35.43}$ & $0.7351^{+0.0500}_{-0.0285}$ & 347.3100 & 0.06881 & $-$2.186 & 15 & M4 &  \\
436214200040092 & 284.6744 & 0.8566 & $^\dagger 455.63^{+114.79}_{-79.21}$ & $0.7328^{+0.2668}_{-0.1047}$ & 348.2973 & 0.09092 & $-$1.931 & 40 & M4 &  \\
437203100058319 & 291.2111 & $-$4.7123 & $^\dagger 2765.85^{+1634.56}_{-1467.85}$ & $1.4917^{+0.1570}_{-0.1570}$ & 348.3126 & 0.21199 & $-$1.911 & 27 & M4 &  \\
437211400092016 & 291.2327 & $-$1.8312 & $^\dagger 3683.40^{+1707.96}_{-1886.50}$ & $0.6804^{+0.1832}$ & 347.3185 & 0.17083 & $-$2.960 & 19 & M3 &  \\
461216200033263 & 99.9443 & 7.4280 & -- & 2.1996 & 853.2143 & 0.09151 & $-$1.848 & 19 & M3 &  \\
462201300001148 & 112.7258 & 1.6316 & $^\dagger 2524.95^{+1756.09}_{-1004.31}$ & $0.3926_{-0.0188}$ & 482.4555 & 0.11739 & $-$2.282 & 27 & M3 &  \\
486211400004409 & 274.1717 & 5.4338 & $^\dagger 3099.40^{+1958.82}_{-1648.63}$ & $0.5496_{-0.0804}$ & 643.4100 & 0.11097 & $-$1.786 & 33 & M4 &  \\
487207400067044 & 280.5736 & 3.2671 & $^\dagger 853.25^{+1527.94}_{-291.60}$ & $2.1523^{+1.0928}_{-0.9774}$ & 644.4040 & 0.06409 & $-$1.737 & 32 & M2 &  \\
488203200156038 & 286.7733 & 2.1147 & $^\dagger 4936.48^{+2652.39}_{-2073.99}$ & $3.1927^{+0.0262}_{-0.1570}$ & 340.2434 & 0.15744 & $-$2.079 & 18 & M0 &  \\
491203400002897 & 308.7275 & 1.5507 & -- & 0.1863 & 670.3563 & 0.10956 & $-$2.056 & 24 & M3 &  \\
536204200026434 & 260.4943 & 8.9845 & -- & 0.2463 & 634.3900 & 0.11730 & $-$1.659 & 33 & M4 &  \\
537204100031453 & 268.7043 & 9.6602 & -- & 0.5060 & 665.4077 & 0.06055 & $-$2.054 & 24 & M4 &  \\
539209100126426 & 288.1225 & 12.9930 & $^\dagger 2767.03^{+616.65}_{-513.43}$ & $3.9255^{+0.1047}_{-0.0785}$ & 334.2062 & 0.07879 & $-$3.418 & 19 & M0 & Present in DR8, absent in DR17 \\
540208400015276 & 289.8905 & 10.4382 & $321.89^{+49.02}_{-37.51}$ & $0.6885^{+0.0383}_{-0.1889}$ & 341.2874 & 0.04033 & $-$2.163 & 23 & M6 & Recurrent flares in $g$- and $r$- bands \\
540215200069194 & 290.4086 & 14.9622 & -- & 11.6013 & 342.2201 & -- & $-$3.039 & 19 & M0 &  \\
542214100014895 & 307.2230 & 15.0669 & $125.11^{+2.53}_{-2.64}$ & 0.0000 & 671.3297 & 0.08716 & $-$2.481 & 29 & M4 &  \\
543206400016038 & 314.4593 & 10.0783 & $^\dagger 1252.52^{+1561.89}_{-690.48}$ & $0.2617_{-0.0523}$ & 672.4002 & 0.16202 & $-$2.763 & 69 & M4 &  \\
543215400016323 & 312.3468 & 13.9935 & -- & 0.2313 & 672.4067 & 0.07693 & $-$2.120 & 19 & M4 &  \\
562216200020648 & 84.6147 & 22.3520 & -- & 2.6877 & 852.2416 & 0.11486 & $-$2.335 & 40 & M1 &  \\
563202400050273 & 96.8058 & 15.3731 & $^\dagger 599.07^{+344.88}_{-187.01}$ & $0.3218^{+0.3324}_{-0.2800}$ & 862.2359 & -- & $-$2.318 & 104 & M3 &  \\
565209300016509 & 112.6538 & 19.2998 & -- & 0.0811 & 795.3952 & 0.10856 & $-$2.170 & 41 & M3 &  \\
588211300040671 & 272.7993 & 19.6388 & -- & 0.2051 & 645.4004 & -- & $-$2.536 & 6 & M4 &  \\
588212300042173 & 271.0911 & 19.3271 & $^\dagger 5856.62^{+2846.60}_{-3822.58}$ & 0.4187 & 645.3718 & -- & $-$1.861 & 3 & M2 &  \\
592201300048015 & 306.0703 & 15.7461 & -- & 0.4941 & 344.1939 & -- & $-$2.713 & 12 & M3 &  \\
611215200019569 & 82.1163 & 29.4668 & -- & 1.5365 & 846.1697 & 0.07061 & $-$2.509 & 33 & M4 &  \\
613214200021207 & 98.8276 & 29.4430 & -- & 0.4955 & 791.4480 & 0.06009 & $-$1.659 & 14 & M4 &  \\
615210400006263 & 115.5509 & 26.5841 & -- & 0.0991 & 849.2762 & 0.15224 & $-$2.549 & 8 & M5 &  \\
615214400005704 & 114.8512 & 28.5188 & $496.64^{+108.08}_{-57.20}$ & $0.0785^{+0.0262}_{-0.0262}$ & 846.3346 & 0.26036 & $-$2.629 & 20 & M4 & Recurrent flares in $r$-band \\
616216400012099 & 118.0565 & 28.6515 & -- & 0.0864 & 812.5463 & -- & $-$3.070 & 6 & M5 &  \\
642215200028716 & 314.0120 & 29.1802 & -- & 0.3274 & 661.3749 & 0.05116 & $-$2.043 & 10 & M4 &  \\
642215300060146 & 314.5377 & 28.5770 & -- & 0.4228 & 802.1148 & 0.26050 & $-$1.797 & 55 & M3 &  \\
655210200003936 & 55.9941 & 34.8996 & $^\dagger 1044.42^{+669.57}_{-358.90}$ & 0.7851 & 789.3148 & 0.12400 & $-$3.808 & 8 & M4 &  \\
660207200039946 & 92.7499 & 32.5936 & -- & 1.3037 & 790.4454 & 0.15593 & $-$2.462 & 44 & M3 &  \\
660207300043882 & 92.0245 & 32.0495 & -- & 1.5532 & 790.4520 & 0.06394 & $-$2.176 & 19 & M4 &  \\
660209300008318 & 97.2417 & 33.7800 & -- & 0.6422 & 790.4604 & 0.18062 & $-$1.904 & 26 & M4 &  \\
684209200042442 & 285.7614 & 35.1056 & $^\dagger 2347.42^{+1476.64}_{-964.59}$ & $0.2617^{+0.0262}$ & 299.2711 & 0.10895 & $-$1.927 & 41 & M4 &  \\
685205100007414 & 294.3103 & 33.0739 & $^\dagger 3403.98^{+1480.39}_{-1627.35}$ & $0.3537^{+0.1986}_{-0.0658}$ & 345.2225 & 0.01509 & $-$3.429 & 15 & M4 &  \\
685211100071699 & 289.4993 & 34.4177 & -- & 0.3306 & 345.2271 & 0.14859 & $-$1.874 & 17 & M3 &  \\
686201100023141 & 302.3502 & 30.7379 & $^\dagger 4042.33^{+2290.52}_{-1536.73}$ & $3.4090^{+0.1061}_{-0.4143}$ & 346.2240 & 0.27842 & $-$1.851 & 26 & M0 &  \\
686208200055661 & 294.3102 & 33.0738 & $^\dagger 3403.98^{+1480.39}_{-1627.35}$ & $0.3537^{+0.1986}_{-0.0658}$ & 345.2221 & 0.10247 & $-$2.593 & 18 & M4 &  \\
687207100049742 & 305.8142 & 33.1314 & $^\dagger 2966.10^{+1774.93}_{-1357.79}$ & $3.0095^{+0.1832}_{-0.5557}$ & 658.3663 & 0.04862 & $-$3.037 & 36 & M0 & Recurrent flares in $r$-band \\
687214100050598 & 307.4067 & 36.2972 & -- & 6.4051 & 658.3508 & 0.16569 & $-$2.569 & 40 & M0 &  \\
688214300032111 & 313.9940 & 35.6929 & $^\dagger 541.88^{+768.49}_{-142.82}$ & $0.2187^{+0.4356}_{-0.1402}$ & 450.1517 & 0.04322 & $-$3.517 & 13 & -- &  \\
689211400045274 & 320.7232 & 33.7768 & $^\dagger 1836.04^{+1467.16}_{-721.42}$ & $0.4187^{+0.0262}_{-0.1832}$ & 449.1489 & 0.18553 & $-$1.887 & 21 & M4 &  \\
690210100033851 & 331.0012 & 34.3635 & -- & 0.3364 & 660.3650 & 0.10866 & $-$3.213 & 47 & M4 &  \\
700213100014818 & 58.7839 & 43.5987 & -- & 1.0205 & 793.3194 & 0.08245 & $-$2.015 & 17 & M4 &  \\
704203100027996 & 88.6826 & 37.8677 & $^\dagger 5060.59^{+5223.01}_{-3068.89}$ & $1.0206_{-0.1832}$ & 812.3948 & 0.05178 & $-$2.974 & 43 & M4 &  \\
706208200005412 & 101.8098 & 39.9440 & -- & 0.3238 & 793.3987 & 0.11954 & $-$2.196 & 22 & M2 &  \\
728205100116115 & 299.7507 & 39.8352 & $^\dagger 7906.58^{+4266.19}_{-2987.71}$ & $0.8374_{-0.0785}$ & 437.0989 & 0.07523 & $-$1.930 & 11 & M2 &  \\
733207400019437 & 337.1025 & 38.7816 & $^\dagger 898.49^{+581.59}_{-194.95}$ & $0.2094^{+0.0262}_{-0.0262}$ & 648.4201 & 0.05229 & $-$2.543 & 49 & M4 &  \\
733209300032227 & 341.4235 & 40.9291 & -- & 0.3876 & 648.4221 & 0.12216 & $-$2.269 & 29 & M4 &  \\
742211400023238 & 61.8394 & 47.9758 & -- & 2.8823 & 806.4472 & 0.06852 & $-$2.324 & 21 & M3 &  \\
766203400032547 & 292.2349 & 44.4207 & $^\dagger 422.97^{+156.86}_{-83.71}$ & $0.0785^{+0.0262}_{-0.0785}$ & 295.4018 & 0.17767 & $-$1.505 & 29 & M4 &  \\
766205100052523 & 297.4471 & 47.3023 & $^\dagger 770.36^{+284.14}_{-203.85}$ & $0.0785^{+0.1006}_{-0.0523}$ & 294.4330 & 0.14369 & $-$1.829 & 22 & M4 &  \\
767206100019391 & 304.8313 & 47.2899 & $^\dagger 4526.71^{+3963.91}_{-2037.80}$ & $1.7534^{+0.0262}_{-0.4972}$ & 448.1220 & 0.14390 & $-$2.514 & 19 & M4 &  \\
767212100038888 & 299.2655 & 49.0555 & -- & 0.3939 & 448.1118 & 0.11812 & $-$1.944 & 11 & M4 &  \\
768202400043820 & 313.9297 & 44.2798 & $178.57^{+32.63}_{-25.55}$ & 0.0000 & 448.1355 & 0.05168 & $-$3.170 & 10 & M6 &  \\
768209200100383 & 316.0178 & 49.1134 & -- & 6.0695 & 451.1700 & 0.04848 & $-$2.940 & 8 & M0 &  \\
768211400063696 & 311.4420 & 48.4437 & $243.37^{+33.22}_{-36.45}$ & $0.0236^{+0.1954}_{-0.0236}$ & 448.1217 & 0.07152 & $-$2.043 & 18 & M6 &  \\
771211400031727 & 341.3488 & 47.8960 & $^\dagger 1981.92^{+1431.51}_{-847.37}$ & $0.4187^{+0.0262}_{-0.0473}$ & 461.1888 & 0.05117 & $-$3.371 & 18 & M4 &  \\
771215100045769 & 341.0910 & 50.7499 & $^\dagger 543.48^{+191.95}_{-143.98}$ & $0.3402^{+0.0173}_{-0.0262}$ & 461.2259 & 0.09286 & $-$2.540 & 29 & M4 &  \\
771216100033044 & 338.0857 & 50.6752 & $450.52^{+38.82}_{-48.54}$ & $0.1570_{-0.0653}$ & 461.2308 & 0.18924 & $-$2.105 & 33 & M3 &  \\
772205100015789 & 357.2015 & 46.8119 & $^\dagger 588.03^{+315.95}_{-132.54}$ & $0.2355_{-0.0262}$ & 649.4396 & 0.11356 & $-$2.326 & 43 & M4 & Recurrent flares in $g$- and $r$- bands \\
772210400025822 & 354.5118 & 48.2560 & -- & 0.4213 & 776.3834 & 0.02708 & $-$3.350 & 21 & M4 &  \\
778208300004589 & 54.8447 & 53.6707 & $461.60^{+71.36}_{-55.11}$ & $1.4394^{+0.0785}$ & 830.1527 & -- & $-$2.071 & 33 & M4 &  \\
800206300002069 & 303.3847 & 53.9294 & $^\dagger 3817.64^{+2514.06}_{-1387.97}$ & $0.7851_{-0.0262}$ & 441.1559 & 0.14065 & $-$2.155 & 42 & M3 & Recurrent flares in $r$-band \\
803205200026342 & 339.1736 & 54.2447 & $^\dagger 3135.60^{+1463.73}_{-1347.46}$ & $1.0168^{+0.3441}_{-0.2840}$ & 468.1810 & 0.12514 & $-$2.141 & 37 & M4 &  \\
803205400072878 & 340.1963 & 53.5281 & $^\dagger 1022.21^{+877.26}_{-347.19}$ & $0.3926^{+0.0262}_{-0.0262}$ & 468.1027 & 0.03316 & $-$1.937 & 25 & M4 &  \\
803215400080106 & 334.9022 & 57.6831 & $^\dagger 2825.96^{+2504.89}_{-1787.94}$ & $2.1770^{+1.3915}_{-0.6684}$ & 468.1069 & 0.13338 & $-$2.120 & 84 & M0 &  \\
804211400018421 & 344.3353 & 55.5434 & $495.73^{+125.58}_{-70.05}$ & $0.5234^{+0.0645}_{-0.1047}$ & 476.1397 & 0.15684 & $-$1.944 & 19 & M4 &  \\
804215300063018 & 342.5648 & 57.6657 & $^\dagger 2464.91^{+1194.40}_{-959.50}$ & $0.9159^{+0.6543}_{-0.0785}$ & 476.1230 & 0.11517 & $-$3.546 & 21 & M4 & Recurrent flares in $r$-band \\
806210400049537 & 9.6717 & 62.3984 & $^\dagger 1423.76^{+636.28}_{-534.20}$ & $1.0468^{+0.0407}_{-0.1579}$ & 473.1159 & 0.13693 & $-$2.836 & 94 & M3 & Flares in both $g$- and $r$- bands \\
807203100058808 & 18.4002 & 60.1114 & $^\dagger 396.42^{+79.30}_{-83.97}$ & $0.7851^{+0.0548}_{-0.0426}$ & 474.1847 & 0.03827 & $-$1.922 & 16 & M4 &  \\
807211100054997 & 19.8124 & 63.7229 & $355.37^{+67.36}_{-51.40}$ & $0.4979^{+0.0800}_{-0.2558}$ & 475.2446 & 0.08254 & $-$2.656 & 15 & M7 &  \\
830208200021745 & 320.1856 & 61.2979 & $^\dagger 937.77^{+362.58}_{-263.98}$ & $1.0438^{+0.1077}_{-0.0493}$ & 777.2401 & 0.12994 & $-$1.724 & 36 & M4 &  \\
831208100003902 & 334.4147 & 61.8522 & $^\dagger 792.92^{+359.06}_{-286.97}$ & $0.9159^{+3.4664}_{-0.2355}$ & 776.2748 & 0.01574 & $-$2.376 & 11 & M3 &  \\
832210400037888 & 356.7705 & 62.2385 & -- & 3.6745 & 775.2605 & 0.19886 & $-$2.038 & 53 & -- &  \\
\bottomrule
\label{tab:Table_1}
\end{longtable}
$^1$ For the objects with defined geometric distance a three-dimensional map of Milky Way dust reddening ``Bayestar19'' \citep{Green_2019} is used, if no ---  we used a map of Galactic Dust Reddening and Extinction by \citealt{2011ApJ...737..103S}. \\
$^2$ Peak time, FWHM, amplitude and number of points are extracted from the parametric fit method. In case of small amount of points, a peak time corresponds to the photometric measurement with the minimum magnitude while an amplitude is calculated as difference between minimal magnitude and magnitude of quiescent star obtained from the parametric fit. FWHM is extracted based on the parametric fit method only for objects with enough points to construct an adequate flare profile.\\
$^\dagger$ Objects with $\text{Plx} / \text{e}\_\text{Plx} < 5$ according to Gaia DR3 parallax estimations.
\end{landscape}

\bsp	
\label{lastpage}
\end{document}